\documentclass[twocolumns]{emulateapj}
\usepackage{natbib}
\usepackage{amssymb}
\usepackage{amsmath}
\def\be{\begin{eqnarray}}
\def\ee{\end{eqnarray}}
\usepackage{comment}
\usepackage[dvipsnames]{xcolor}
\usepackage{graphicx,subfigure}
\usepackage{enumerate}
\usepackage{rotating}
\usepackage{longtable}
\usepackage{booktabs}
\usepackage{float}

\usepackage{hyperref}
\hypersetup{colorlinks,urlcolor={blue},linkcolor={blue},citecolor={blue}}

\shorttitle{Radio follow-up of GW170817}

\shortauthors{}

\begin{document}

\title{Continued radio observations of GW170817 3.5 years post-merger}

\author{Arvind~Balasubramanian\altaffilmark{1}}
\author{Alessandra~Corsi\altaffilmark{1}}
\author{Kunal~P.~Mooley\altaffilmark{2}}
\author{Murray~Brightman\altaffilmark{2}}
\author{Gregg~Hallinan\altaffilmark{2}}
\author{Kenta~Hotokezaka\altaffilmark{3}}
\author{David~L.~Kaplan\altaffilmark{4}}
\author{Davide~Lazzati\altaffilmark{5}}
\author{Eric~J.\ Murphy\altaffilmark{6}}
\altaffiltext{1}{Department of Physics and Astronomy, Texas Tech University, Box 1051, Lubbock, TX 79409-1051, USA; e-mail: arvind.balasubramanian@ttu.edu}
\altaffiltext{2}{Caltech, 1200 E. California Blvd. MC 249-17, Pasadena, CA 91125, USA}
\altaffiltext{3}{Research Center for the Early Universe, Graduate School of Science, University of Tokyo, Bunkyo-ku, Tokyo 113-0033, Japan}
\altaffiltext{4}{Center for Gravitation, Cosmology, and Astrophysics,
Dept. of Physics, University of Wisconsin-Milwaukee, P.O. Box 413, Milwaukee, WI 53201, USA}
\altaffiltext{5}{Department of Physics, Oregon State University, 301 Weniger Hall, Corvallis, OR 97331, USA}
\altaffiltext{6}{National Radio Astronomy Observatory, Charlottesville, VA 22903, USA}

\begin{abstract}
 We present new radio observations of the binary neutron star merger GW170817 carried out with the Karl G. Jansky Very large Array (VLA) more than 3\,yrs after the merger.  Our combined dataset is derived by co-adding more than $\approx32$\,hours of VLA time on-source, and as such provides the deepest combined observation (rms sensitivity $\approx 0.99\,\mu$Jy) of the GW170817 field obtained to date at 3\,GHz. We find no evidence for a late-time radio re-brightening at a mean epoch of $t\approx 1200$\,d since merger, in contrast to a $\approx 2.1\,\sigma$ excess observed at X-ray wavelengths at the same mean epoch. Our measurements agree with expectations from the post-peak decay of the radio afterglow of the GW170817 structured jet. Using these results, we constrain the parameter space of models that predict a late-time radio re-brightening possibly arising from the high-velocity tail of the GW170817 kilonova ejecta, which would dominate the radio and X-ray emission years after the merger (once the structured jet afterglow fades below detection level). Our results point to a steep energy-speed distribution of the kilonova ejecta (with energy-velocity power law index $\alpha \gtrsim 5$). We suggest possible implications of our radio analysis, when combined with the recent tentative evidence for a late-time re-brightening in the X-rays, and highlight the need for continued radio-to-X-ray monitoring to test different scenarios.
\end{abstract}


\section{Introduction}\label{sec:intro}
GW170817 has been a milestone event for transient multi-messenger studies. It was the first binary neutron star (NS) merger observed by the LIGO and VIRGO detectors \citep{2017PhRvL.119p1101A}, and so far it remains the only binary NS system from which gravitational waves (GWs) and a multi-wavelength (radio to gamma-ray) counterpart have been discovered \citep{2020arXiv201014527A,2020ApJ...905..145K,2021ApJ...912..128P}. The GW170817 NS-NS merger occurred at 12:41:04 on 2017 August 17 UTC, and its GW detection was followed by the detection of a $\gamma$-ray burst (GRB) by the \emph{Fermi} and \emph{INTEGRAL} satellites, $\approx2$\,s after the merger. UV/optical/IR instruments subsequently identified the so-called kilonova counterpart (AT2017gfo), in the galaxy NGC\,4993 at a distance of $\approx40$\,Mpc, making GW170817/GRB170817a the closest short GRB with known redshift \citep[e.g.,][]{2017Natur.551...64A,2017ApJ...848L..19C,2017Sci...358.1556C,2017ApJ...848L..17C,2017Sci...358.1570D,2017Sci...358.1559K,2017Natur.551...80K,2017Sci...358.1583K,2017Natur.551...67P,2017Sci...358.1574S,2017Natur.551...75S,2017ApJ...848L..27T,2017ApJ...848L..24V}. Observations of the quasi-thermal UV/optical/IR emission from the GW170817 slow ($\sim 0.1c-0.3c$), neutron-rich, kilonova ejecta were successful in verifying that mergers of NSs in binaries are production sites of heavy elements such as gold and platinum \citep[e.g.,][]{2017Sci...358.1559K,2017Natur.551...80K,2017Natur.551...67P,2017arXiv171005931M}. 

In addition to the quasi-thermal kilonova emission, a delayed non-thermal (synchrotron) afterglow from GW170817/GRB\,170817a was first observed in the X-rays $\approx 9$\,days after the merger by the \emph{Chandra} observatory \citep[e.g.,][]{2017Natur.551...71T,2017ApJ...848L..25H,2017ApJ...848L..20M}. A radio afterglow detection with the Karl G. Jansky Very Large Array (VLA) followed, about two weeks after the merger  \citep{2017Sci...358.1579H}. Further radio observations of the source proved decisive in narrowing down the morphology of the jet \citep{2018ApJ...861L..10C,2018ApJ...858L..15D,2017ApJ...848L..21A,2018ApJ...856L..18M,2018ApJ...868L..11M, 2018Natur.554..207M,2019ApJ...886L..17H}, ruling out the simple uniform energy-velocity (top-hat) ejecta in favour of a structured jet, where the ejecta velocity varies with the angle from the jet axis \citep{2018Natur.561..355M,2018PhRvL.120x1103L,2020ApJ...901L..26R,2019Sci...363..968G}. These observations, with the help of hydrodynamic simulations \citep{2018PhRvL.120x1103L,2018ApJ...867...18N}, set constraints on the opening angle of the jet core ($\lesssim 5$\,deg), the observer's viewing angle ($\approx 15-30$\,deg), the isotropic equivalent energy ($\sim10^{52}$\,erg) and the interstellar medium (ISM) density ($\sim 10^{-4}-0.5$\,cm$^{-3}$). 

The extended radio follow-up of GW170817 up to 2.1\,years after the merger had shown that the radio emission from the structured jet had faded below typical flux density sensitivities that can be reached with the VLA in a few hours of observing \citep{2020arXiv200602382M}. Several theoretical scenarios, however, predict the possible emergence at late times of detectable electromagnetic emission associated with the afterglow of the kilonova ejecta itself \citep[e.g.,][]{2011Natur.478...82N,2013MNRAS.430.2121P,2015MNRAS.450.1430H,2019MNRAS.487.3914K,2018ApJ...867...95H,2020MNRAS.495.4981M}. Indeed, numerical simulations show that during the NS-NS merger, a modest fraction of a solar mass is ejected from the system, and the total ejecta mass and velocity distribution of such ejecta depend on the total mass, mass ratio, and the nuclear equation of state (EoS) of the compact objects in the binary. While optical-UV observations are mostly sensitive to the low-end of the ejecta velocity distribution, radio (and X-rays) can probe the fastest moving ejecta tail, shedding light on whether NS-NS ejecta are broadly distributed in energy and velocity (as simulations seem to suggest) and providing indirect constraints on the nuclear EoS. 

Tentative evidence for a very late-time re-brightening possibly associated with the kilonova afterglow of GW170817 has recently been reported in the X-rays \citep{2020GCN.29055....1H,2021GCN.29375....1H,2020GCN.29038....1T}. On the other hand, relatively shallow observations (rms $\approx4.3\,\mu$Jy) with the VLA at 3\,GHz had reported a lack of radio detection contemporaneous with the X-ray late-time re-brightening \citep{2020GCN.29053....1A}. This radio non-detection was interpreted to be compatible with expectations from the simplest extrapolation of the X-ray excess to the radio band \citep{2020GCN.29053....1A}.  Here, we use much deeper VLA observations to show that the lack of a detection of radio emission in excess to that expected from the structured jet associated with GW170817 constrains kilonova ejecta models to a relatively steep energy-velocity ejecta profile.

Our work is organized as follows. We report our new observations in $\S$\ref{sec:obs}; in $\S$\ref{sec:disc} we discuss our results within the kilonova afterglow model; finally, in $\S$\ref{sec:conc} we conclude with a summary.  


\setlength\LTcapwidth{\linewidth}
\begin{longtable*}{cccccccc}
\caption{VLA late-time observations of the GW170817 field. See text for details on rms measurements.
\label{tab:rad_obs2}}\\
\toprule
\toprule
Date  & $\nu$  & VLA & Time on-source & rms & VLA   & PI & Nominal synth. beam\\
(UT) & (GHz) & config. & (hr) & ($\mu$Jy) & program & &(\arcsec)\\
\midrule
2020 Sep 19 & 3.0 & B & 2\,h43\,m24\,s & 4.6 & 20A-185 &  Balasubramanian & 2.1\\
2019 Sep 20 & 3.0 & B & 2\,h43\,m27\,s & 5.8 & 20A-185 &  Balasubramanian & 2.1\\
2020 Dec 15 & 3.0 & A & 3\,h24\,m14\,s & 3.5 & SL0449 &  Margutti & 0.65\\
2020 Dec 27 & 3.0 & A & 3\,h24\,m14\,s & 3.3 & SL0449 & Margutti & 0.65\\
2021 Jan 10 & 3.0 & A & 2\,h41\,m34\,s & 3.7 & 20B-208 & Balasubramanian & 0.65\\
2021 Jan 16 & 3.0 & A & 2\,h38\,m34\,s & 3.6 & 20B-208 &  Balasubramanian & 0.65\\
2021 Feb 02 & 3.0 & A & 3\,h24\,m16\,s & 3.3 & SM0329 &  Margutti& 0.65 \\
2021 Feb 04 & 3.0 & A & 2\,h38\,m38\,s & 4.1 & 20B-472 &  Corsi & 0.65\\
2021 Feb 05 & 3.0 & A & 2\,h41\,m41\,s & 4.0 & 20B-472 & Corsi & 0.65\\
2021 Feb 06 & 3.0 & A & 2\,h41\,m34\,s & 3.8 & 20B-472 & Corsi& 0.65 \\
2021 Feb 08 & 3.0 & A & 2\,h41\,m36\,s & 4.0 & 20B-472 & Corsi & 0.65\\
2021 Feb 10 & 15.0 & A & 2\,h40\,m52\,s & 1.9 & SM0329 & Margutti& 0.13 \\
\bottomrule
\end{longtable*}

\section{Observations and data reduction}\label{sec:obs}
As we describe in what follows, we have processed new and archival data of GW170817 obtained at radio and X-ray wavelengths between 2020 September and 2021 February.
The full panchromatic afterglow light curve of GW170817 is available for download on the web\footnote{\url{https://github.com/kmooley/GW170817/} or \url{http://www.tauceti.caltech.edu/kunal/gw170817/}}.

\subsection{Radio Observations}
Radio observations of the GW170817  field were carried out with the VLA  on the dates listed in Table \ref{tab:rad_obs2} at S band (2--4 GHz, nominal central frequency of 3\,GHz) with the array in its B (September 2020) and A configurations (December 2020 -  February 2021). Each observation was calibrated in \texttt{CASA} \citep{McMullin2007} using the automated VLA calibration pipeline. The calibrated data were then manually inspected for further radio frequency interference (RFI) excision. We interactively imaged each observation using the \texttt{CASA} task \texttt{tclean} with one Taylor term (\texttt{nterms=1}) and robust weighting (\texttt{robust=0.5}), and derived the rms measurements using \texttt{imstat}. Table \ref{tab:rad_obs2} lists the  rms sensitivity reached in each observation, estimated within a region of 20 synthesized beams around the position of GW170817 \citep[$\alpha=13{\rm h}09{\rm m}48.069{\rm s}$, $\delta=-23{\rm d}22{\rm m}53.39{\rm s}$, J2000;][]{2018Natur.561..355M}. We find no significant ($>3\times$rms) excess in a region of one synthesized beam around the position of GW170817 in any of the individual images. Next, we co-added and interactively imaged as above (\texttt{nterms=1} and \texttt{robust=0.5}) all A-configuration 3\,GHz VLA observations listed in Table \ref{tab:rad_obs2}. An rms of 1.3\,$\mu$Jy was reached at 2.8\,GHz (Table \ref{tab:obs}) within a region of size approximately equal to 20 synthesized beams centered on the position of GW170817. Within one synthesized beam centered on the location of GW170817, we measure an \texttt{imstat} peak flux density value of  $\approx2.8\,\mu$Jy at 2.8\,GHz. With this procedure, several of the bright, extended sources present in the field left substantial deconvolution residuals. Thus, to mitigate the effects of deconvolution residuals, test the robustness of our measurement,  and further improve our sensitivity, we imaged all 3\,GHz data (both A and B configurations of the VLA) listed in Table \ref{tab:rad_obs2} non-interactively with two Taylor terms (\texttt{nterms=2}), robust weighting (\texttt{robust=0.5}), single phase-only selfcal (solution interval of 4 minutes), and a cleaning threshold of 4\,$\mu$Jy. This yielded an  image rms noise of 0.99\,$\mu$Jy (in a region of size equal to 20 synthesized beams around the position of GW170817; theoretical thermal noise $\approx 0.85\,\mu$Jy) and a peak flux density value of 2.86\,$\mu$Jy (Table \ref{tab:obs}) within one synthesized beam centered on the location of GW170817. 

A late-time VLA observation of the GW170817 field was also carried out in U band (nominal central frequency of 15\,GHz) on 10 February 2021 (Table \ref{tab:rad_obs2}). We calibrated this dataset and interactively imaged the field (with \texttt{nterms=1} and \texttt{robust=0.5}). No significant emission is found at the location of GW170817 (Table \ref{tab:obs}). The rms measured in a region of size equal to 20 synthesized beams centered around the position of GW170817 is of $\approx 1.9\,\mu$Jy at 15\,GHz. 

\begin{figure*}
    \centering
    \includegraphics[width=0.9\textwidth]{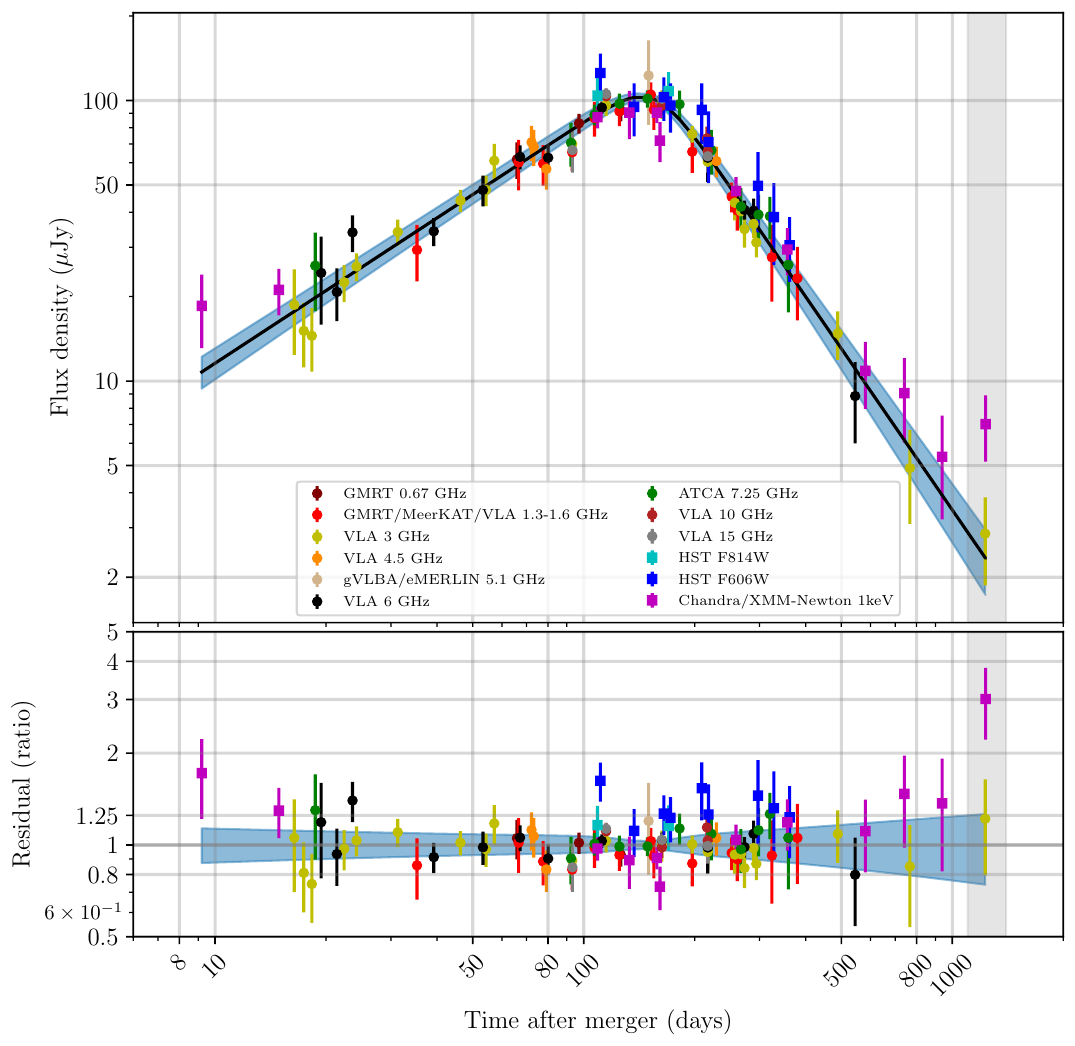}%
    \caption{Comprehensive 3\,GHz light curve of GW170817 as presented in our recent work \cite{2020arXiv200602382M}, which includes data from \citet{2019ApJ...883L...1F,2019Sci...363..968G,2018ApJ...862L..19N}, together with our latest measurement in the radio (3\,GHz, latest yellow data point in the grey, shaded region) and X-rays (latest purple data point in the grey, shaded region) extrapolated to 3\,GHz using the spectral index derived in \cite{2020arXiv200602382M}. The best fit structured jet model for GW170817 is also plotted (top panel, black line) along with the associated $1\sigma$ error region (blue shaded region). As evident from the lower panel, our radio measurement is compatible with the tail of the GW170817 jet within the large errors. On the other hand, the X-rays show a $\sim2\,\sigma$ excess and could indicate the onset of a new component \citep[][]{2020GCN.29055....1H,2021GCN.29375....1H,2020GCN.29038....1T}.}\label{fig:panchrom_lc}%
\end{figure*}

\subsection{X-ray Data}

We reprocessed and analyzed the Chandra ACIS-S observations of the GW170817 field obtained between 2020 December 10 and 2021 January 27 (obsIDs 22677, 24887, 24888, 24889, 23870, 24923, 24924; 150.5 ks, PI Margutti) using the same procedure described in \cite{2020arXiv200602382M}. We find an unabsorbed flux density of $(1.70\pm0.45)\times10^{-4} \,\mu$Jy at $2.4\times10^{17}$\,Hz (1\,keV; $1\sigma$ uncertainty) by combining the spectral products of all seven observations and fitting the data with an absorbed power-law model where the hydrogen column density $N_{\rm H}$ has been fixed to the Galactic value, and where the photon index $\Gamma$ ($N_{E_{\nu}}\propto E_{\nu}^{-\Gamma}$,where $E_{\nu}=h\nu$) has been fixed to 1.58 \citep{2020arXiv200602382M}. To investigate if $\Gamma$ is different from 1.58, we refitted the Chandra data leaving $\Gamma$ as a free parameter. From the 2020 December--January 2021 data we find $\Gamma=2.71^{+1.43}_{-1.08}$. If we additionally combine the 96.6\,ks of data obtained in 2020 March, we get $\Gamma= 1.84^{+0.80}_{-0.73}$ (90\% uncertainties). Hence, in both cases, the value of $\Gamma$ is consistent (well within the 90\% confidence interval) with $\Gamma=1.58$.
Our results are also consistent with \citet[][]{2020GCN.29055....1H,2021GCN.29375....1H,2020GCN.29038....1T,2021arXiv210413378T}.


\setlength\LTcapwidth{\linewidth}
\begin{longtable*}{ccccccc}
\caption{Results for the co-added late-time radio observations of GW170817. See text for discussion.
\label{tab:obs}}\\
\toprule
\toprule
Date  & Epoch & $\nu$ & $F_{\nu}$ & $\sigma_{\nu}$ & Instrument & Reference\\
(UT) & (days) & (Hz) & ($\mu$Jy) & ($\mu$Jy)& &\\
\midrule
2020 Dec 15 - 2021 Feb 08 & 1243 & $2.8\times10^9$ & 2.8 & 1.1-1.3 & VLA A & This work\\
2020 Sep 19 - 2021 Feb 08 & 1199 & $3.0\times10^9$ & 2.86 & 0.99 & VLA A\&B & This work\\
2020 Dec 10 - 2021 Jan 27 & 1234 & $2.41\times10^{17}$ & $1.70\times10^{-4}$ & $0.45\times10^{-4}$ & Chandra& This work\\
\bottomrule
\end{longtable*}

\begin{figure}
\centering
\includegraphics[width=0.48\textwidth]{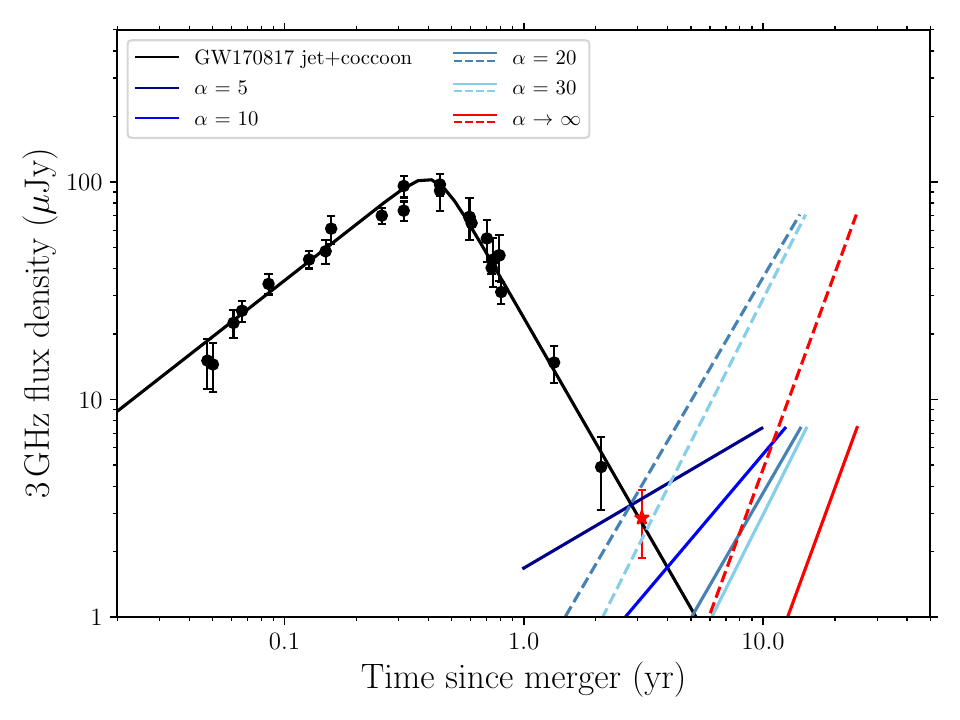}
\vspace {-0.5cm}
\caption{3\,GHz radio light curve of GW170817 with our recent radio measurement (red star) along with predictions for the rising part of the kilonova afterglow light curve as a function of $\alpha$ (see $\S$\ref{eq:kn_rise}) with the assumption that the minimum speed of the ejecta is $\beta_{0}=0.3$. The solid lines assume $p=2.1$, $\epsilon_{e}=7.8\times10^{-3}$, $\epsilon_{\rm B}=9.9\times10^{-4}$, $n=9.8\times10^{-3}{\rm cm}^{-3}$ \citep[as in][]{2020arXiv200602382M} and $\alpha=5, 10, 20, 30, \infty$. These values of $\alpha$ are all compatible with our latest radio measurement (red star), while smaller values of $\alpha$ would produce radio emission in excess to it. For comparison, the dashed lines show the case $\epsilon_{e}=10^{-1}$, $\epsilon_{\rm B}=10^{-3}$, $n=10^{-2}{\rm cm}^{-3}$, $p=2.2$ \citep[as in][]{2019MNRAS.487.3914K}, with  $\alpha=20, 30, \infty$ (see $\S$\ref{sec:disc}.)
\label{fig:kn_lc}}
\end{figure}

\section{Discussion}\label{sec:disc}
As is evident from Figure \ref{fig:panchrom_lc}, our late-time radio observations of GW170817 do not provide evidence for radio emission in excess to what is expected from the very late time tail of a structured jet afterglow model (black solid line). The X-ray observations, on the other hand, suggest a more pronounced statistical fluctuation or the possible emergence of a new component at higher frequencies (bottom panel in Figure \ref{fig:panchrom_lc}). 

The radio-to-X-ray spectral index as derived from the measured ratio of the late-time radio-to-X-ray flux densities (see Table \ref{tab:obs}) is $\Gamma_{\rm radio-X}+1=-0.535\pm0.024$, within $\approx 2.1\sigma$ of the value adopted in Figure \ref{fig:panchrom_lc} ($-0.584\pm0.002$) derived by \cite{2020arXiv200602382M}. The current measurements still carry too large uncertainties for claiming any clear evidence for a change in spectral behavior at late times. It may be possible however that the non-detection of a radio rebrightening associated with a kilonova afterglow, together with the tentative X-ray excess, suggest a flattening of the radio-to-X-ray spectrum at late times.

In general terms, it is not difficult to envision a scenario in which the electron index ($p$) for the ejecta responsible for the late-time X-ray excess differs from the one used to model the structured jet afterglow at earlier times \citep[$p=2.07-2.14$;][]{2020arXiv200602382M}. The predictions of Fermi particle acceleration imply that the  power-law  index  $p$ expected  at  non-relativistic  shock speeds is close to $p\approx 2$, while at ultra-relativistic velocities one can have $p\approx 2.2$ \citep[e.g.][and references therein]{2015SSRv..191..519S}. Thus, a flattening of the radio-to-X-ray spectral index in GW170817, if confirmed by further follow-up, could support the idea that a non-relativistic ejecta component is starting to dominate the emission. 

Differently from the above theoretical predictions for particle acceleration, non-relativistic ejecta observed in radio-emitting core-collapse supernovae typically have $p=2.5-3.2$ \citep[e.g.,][]{1998ApJ...499..810C}, pointing to steeper radio-to-X-ray spectra than that suggested by the X-ray excess observed in GW170817. Cases where a transition of the ejecta from the relativistic to the non-relativistic regime has been observed include  GRB\,030329 \citep[e.g.,][]{frail2005,2008A&A...480...35V} and TDE Swift J1644+57 \citep{2021ApJ...908..125C}. These two cases pointed to a slowly-increasing or constant value of $p$ in the relativistic-to-non-relativistic transition, which again would correspond to a spectral steepening rather than a spectral flattening. 

One key difference between GW170817 and other non-relativistic flows is that
we have seen a single power-law spectrum from radio, optical, and X-rays,  i.e. 
synchrotron radiation from electrons with different Lorentz factors in many orders of magnitude. 
In this case, the spectral index should represent $p$. On the other hand, in other non-relativistic flows, we often determine $p$ from the radio spectra closer to the minimum Lorentz factor $\gamma_m$ of the electron energy distribution,  where the synchrotron emission may be dominated by thermal electrons around the typical Lorentz factor rather than accelerated electrons \citep[e.g.,][]{2015PhRvL.114h5003P,2013ApJ...762L..24M}. Thus, GW170817 offers an opportunity to test particle acceleration theory, and continued monitoring from radio-to-X-rays is key to this end. We also note that continued X-ray observations may offer an opportunity to probe the evolution of the cooling frequency in the Newtonian limit, when $\nu_{\rm c}\propto\beta^{-3}t^{-2}$ \citep{2018ApJ...867...95H}, and thus constrain $\beta$ (velocity of the ejecta in units of $\boldsymbol{c}$) and the kinetic energy of the fast tail of the ejecta \citep[see also][]{2019MNRAS.483..624L}. 

An alternative explanation for the excess in X-rays (as compared to the radio) observed in GW170817 might be the possibility of a Compton echo of the X-rays from the prompt emission of  GRB\,170817A, scattering off surrounding dust \citep{2018MNRAS.476.5621B}. Given currently large uncertainties in the X-ray result, hereafter we focus on the constraints that the lack of a radio excess set on kilonova ejecta models. 

Following \cite{2019MNRAS.487.3914K}, the kilonova blast wave drives a shock through the interstellar medium, resulting in synchrotron emission. Electrons are accelerated to a power-law distribution of Lorentz gamma factors $\gamma_{e} > \gamma_{e,m}$, with power-law index $p$. The energy in the kilonova blast wave is distributed as $E(>\beta \gamma)\propto(\beta\gamma)^{-\alpha}$ (with $\gamma$ the Lorentz factor of the shocked fluid) and normalized to  the  total energy  $E$  at  some  minimum  velocity $\beta_0$ such that $E>(\beta_0\gamma_0)=E$. It is reasonable to assume that radio (GHz) observations are in between the minimum frequency, $\nu_{m}$ \citep[corresponding to $\gamma_{m}$, see][] {2011Natur.478...82N}, and the cooling frequency, $\nu_{c}$.  In this case, the kilonova peak flux density reads \citep{2011Natur.478...82N}:  
\begin{equation}
    F_{\rm \nu, pk}\approx (1522\,{\rm \mu Jy})\,\epsilon_{\rm e,-1}^{p-1}\,\epsilon_{\rm B,-3}^{\frac{p+1}{4}}\,n_{-2}^{\frac{p+1}{4}}\,\beta_{0}^{\frac{5p-7}{2}}\,E_{51}\,\nu_{9.5}^{\frac{1-p}{2}}\,d_{26}^{-2},
\end{equation}
where $Q_{x}=Q/10^{x}$ is followed for all quantities ($Q$, all expressed in cgs units); $\epsilon_{\rm B}$ and $\epsilon_{\rm e}$ are the fractions of the total energy in the magnetic field and electrons respectively; $n$, the number density of the medium; $d$ is the distance to the source; the normalization  constant is calculated for $p=2.1$. The time at which the kilonova afterglow emission peaks can be calculated as \citep{2019MNRAS.487.3914K}:
\begin{equation}
    t_{\rm dec}=t_{\rm pk}\approx (3.3 {\rm yr})\left(\frac{E_{\rm iso, 51}}{n_{-2}}\right)^{\frac{1}{3}}\, \beta_{0}^{-\frac{2}{3}}\left(\frac{2+\alpha}{\beta_0(5+\alpha)}-1\right).
\end{equation}

The blast wave can be approximated to be mildly relativistic before this peak, and therefore the rising part of the kilonova ejecta light curve can be easily modeled as \citep[see][and references therein]{2019MNRAS.487.3914K}:
\begin{equation}\label{eq:kn_rise}
    F_{\rm\nu, KN}\,(t)=F_{\rm \nu,pk}\left(\frac{t}{t_{\rm p}}\right)^s,
\end{equation}
where:
\begin{equation}
    s=\frac{3\alpha-6(p-1)}{8+\alpha}.
\end{equation}

In Figure \ref{fig:kn_lc}, we plot the rising portion of the 3\,GHz kilonova light curves obtained following the above prescriptions, and setting $\beta_{0}=0.3$. This choice is motivated by the fact that observations in UV/optical/IR of the early kilonova, which only probe the slowest-moving material, point to speeds of $\sim 0.1c-0.3c$ (see Section \ref{sec:intro}). Since we expect the radio to probe the fastest tail of the kilonova ejecta, we consider $\beta_0\sim 0.3$ a reasonable choice. We note however that smaller values of $\beta_0$, though unlikely, would shift the radio light curve peak to later times, thus allowing for less steep values of $\alpha$. We set $E_{\text{iso}}=10^{51}$\,erg, $d = 40$\,Mpc, while varying the power-law index $\alpha$ of the energy-speed distribution of the kilonova ejecta. The solid lines correspond to the choice $p=2.1$, $\epsilon_{e}=7.8\times10^{-3}$, $\epsilon_{\rm B}=9.9\times10^{-4}$\, $n=9.8\times10^{-3}{\rm cm}^{-3}$ as derived from the modeling of the earlier-time panchromatic afterglow of the GW170817 structured jet \citep{2020arXiv200602382M}. For comparison, the dashed lines show the case $\epsilon_{e}=10^{-1}$, $\epsilon_{\rm B}=10^{-3}$\, $n=10^{-2}{\rm cm}^{-3}$, $p=2.2$, which corresponds to the generic case discussed in \cite{2019MNRAS.487.3914K}. As evident from this Figure, to explain the absence of a kilonova detection in the radio one needs $\alpha\gtrsim5$ for the case where the density and micro-physical parameters are set equal to the ones measured for the structured jet afterglow. This constraint on $\alpha$ agrees with the predictions from numerical simulations described in \cite{2018ApJ...867...95H} and X-ray observations discussed in \cite{2020RNAAS...4...68H}. For the more generic parameters as in \cite{2019MNRAS.487.3914K}, $\alpha\gtrsim20$. 

For an equal mass ratio binary, a steeper energy-velocity distribution at a given $\beta \gamma$ correlates with a stiffer NS EoS for a given cold, non-rotating maximum mass \citep[compare e.g. $SFHo$ and $LS220$ in Figures 1 and 9 of][]{2018ApJ...869..130R}. For EoS with the same stiffness (i.e. with the same radii at NS masses of $1.4$\,M$_\odot$),  larger values of the cold, non-rotating maximum NS mass also correlate to steeper energy-velocity ejecta distribution \citep[as long as a NS is formed even if for a short timescale of order $\sim 1$\,ms; compare e.g. $BHB\Lambda\phi$ and $DD2$ in Fig. 9 of][]{2018ApJ...869..130R}. On the other hand, if a fast tail exists in the ejecta, one can robustly exclude stiff EoS and relatively high mass ratio scenarios due to the weak or absent core bounce in these scenarios \citep[][]{2021arXiv210404537N}. Taken together, if future radio observations reveal a kilonova afterglow, these trends would favor moderate stiffness and mass ratio models. Given these considerations, the constraints we are setting here on $\alpha$ shed some light on the possible EoS, but cannot uniquely pinpoint it. An independent measurement of $\beta$ via direct size imaging (once the ejecta becomes bright enough in the radio) together with constraints on $\alpha$ derived from  light curve modeling, may help reducing degeneracies.

\section{Summary and conclusion}\label{sec:conc}
We have presented extensive late-time radio observations of the GW170817 field, carried out with the most extended configurations of the VLA. Combining the collected data we have built the deepest high-resolution image of the GW170817 field available so far. Our radio flux density measurements show that there is no evidence for emission in excess to the one expected from the afterglow of the GW170817 structured jet at 3\,GHz and $t\approx 1200$\,days since merger. These results constrain the energy-speed distribution of the kilonova ejecta to be rather steep, with a power-law index of $\alpha \gtrsim 5$ (for $\beta_0\gtrsim 0.3$). We finally commented on how the recent detection of a potential excess in the X-rays may hint to a flattening of the power-law index (albeit $\approx2.1\,\sigma$ in terms of significance) of the electron energy distribution of the kilonova ejecta compared to the value of this parameter as constrained by the earlier panchromatic afterglow observations. Further late-time monitoring of the GW170817 field with the VLA is likely to reveal whether a kilonova afterglow is emerging. 

\acknowledgements
We are grateful to Dale A. Frail  for contributing to shaping this work with many insightful discussions. A.B. and A.C. acknowledge support from the National Science Foundation via Award \#1907975. K.P.M. and G.H. acknowledge support from the National Science Foundation Grant AST-1911199. D.L.K. was supported by NSF grant AST-1816492. D.L. acknowledges support from the National Science Foundation via Award \#1907955. The National Radio Astronomy Observatory is a facility of the National Science Foundation operated under cooperative agreement by Associated Universities, Inc. 
\bibliography{references}
\end{document}